\documentstyle[prl,aps,floats,twocolumn]{revtex}
\input{psfig.tex}

\newcommand{\newc}{\newcommand}
\newc\eg{{\it {e.g.}}}

\newcommand\fa{f_{a}}
\newcommand\mchi{m_{\chi}}              \newcommand\nchi{n_{\chi}}

\newcommand\photino{\widetilde{\gamma}} \newcommand\mphotino{m_{\photino}}

\newcommand\axino{\widetilde{a}}        \newcommand\maxino{m_{\axino}}
\newcommand\abunda{\Omega_{\axino}h^2}

\newc{\cachigamma}{C_{a\chi\gamma}}

\newc{\caww}{C_{aWW}}                   \newc{\cayy}{C_{aYY}}
\newc{\sthw}{\sin\theta_W}              \newc{\cthw}{\cos\theta_W}
\newc{\bino}{\widetilde B}              \newc{\wino}{\widetilde W_3}
\newc{\higgsinob}{{\widetilde H}^0_b}   \newc{\higgsinot}{{\widetilde H}^0_t}

\newc{\abund}{\Omega h^2}
\newc{\abundchi}{\Omega_\chi h^2}
\newc{\rhocrit}{\rho_{crit}}
\newc{\rhochi}{\rho_{\chi}}

\newc{\mplanck}{M_{\rm P}}
\newc{\xf}{x_f}
\newc{\jxf}{J({\xf})}
\newc{\VEV}[1]{\langle #1 \rangle}

\newcommand\tev{\,\mbox{TeV}}
\newcommand\gev{\,\mbox{GeV}}
\newcommand\mev{\,\mbox{MeV}}
\newcommand\kev{\,\mbox{keV}}
\newcommand\ev{\,\mbox{eV}}

\newc{\ra}{\rightarrow}
\newc{\beq}{\begin{equation}}
\newc{\eeq}{\end{equation}}
\newc{\bea}{\begin{eqnarray}}
\newc{\eea}{\end{eqnarray}}

\renewcommand\({\left(}
\renewcommand\){\right)}
\renewcommand\[{\left[}
\renewcommand\]{\right]}

\newcommand\lsim{\mathrel{\rlap{\lower4pt\hbox{\hskip1pt$\sim$}}
    \raise1pt\hbox{$<$}}}
\newcommand\gsim{\mathrel{\rlap{\lower4pt\hbox{\hskip1pt$\sim$}}
    \raise1pt\hbox{$>$}}}

\begin{document}
\draft
\preprint{LANCS-TH/9824, SNUTP 98-144, KIAS-P98048 
}       

\title{Axinos as Cold Dark Matter}

\author{Laura Covi$^{1}$, Jihn E. Kim$^{2,3}$, and Leszek Roszkowski$^{1}$}

\address{$^{(1)}${\it Department of Physics, Lancaster University, 
Lancaster LA1 4YB, England}}

\address{$^{(2)}${\it Center for Theoretical Physics, Seoul
National University, Seoul 151-742, Korea}}

\address{$^{(3)}${\it Korea Institute for Advanced Study,
Cheongryangri-dong, Seoul 130-012, Korea}}

\date{\today}
\maketitle

\begin{abstract}

We show that axinos produced in the early Universe in the decay of the
lightest neutralinos are a natural candidate for {\em cold} dark
matter.  We argue that axinos may well provide the main component of
the missing mass in the Universe because their relic density is often
comparable with the critical density.

\end{abstract}




\noindent
{\it 1. Introduction}.\
Axinos are predicted to exist in models involving low-energy
supersymmetry (SUSY) and the Peccei-Quinn solution~\cite{pq} to the
strong CP problem. They are supersymmetric partners of
axions~\cite{axion,kim,dfsz}.  SUSY is widely considered as perhaps the
most attractive framework in which the Fermi scale can be naturally
connected with physics around the Planck scale. The Peccei-Quinn (PQ)
mechanism, which invokes a global, chiral $U(1)$ symmetry group
spontaneously broken at some high energy scale $\fa\sim10^{11}\gev$ 
remains the most compelling way of solving the strong CP problem.

Axinos are thus very strongly motivated. Despite this, they have
received much less attention in the literature than other SUSY
partners. Of particular importance to both experimental searches and
cosmology is the lightest supersymmetric particle (LSP). Axinos, being
massive and electrically and color neutral are an interesting
candidate for the LSP.  One of the most important consequences of
supersymmetry for cosmology in the presence of unbroken R-parity is
the fact that the LSP is stable and may contribute substantially to
the relic mass density in the Universe. If the contribution is of
order the critical density $\rhocrit$, such a particle is considered
an attractive dark matter (DM) candidate. Current models of 
the formation of large structures as well as measured
shape of their power spectrum strongly suggest that a dominant
contribution to the dynamical component of the total mass-energy
density is that from cold DM~\cite{kt}.

In the minimal SUSY model (MSSM), the LSP is usually {\em assumed} to
be the lightest of the four neutralinos. The (lightest) neutralino
$\chi$ is a mixture $\chi= Z_{11}\bino + Z_{12}\wino +
Z_{13}\higgsinob + Z_{14}\higgsinot$ of the respective fermionic
partners (denoted by a tilde) of the electrically neutral gauge bosons 
$B$ and $W_3$, and Higgs bosons $H_b$ and $H_u$.  
It is well-known that the
neutralino's relic density $\rhochi$ is often of order $\rhocrit$. At
high temperatures in the early Universe a thermal population of
neutralinos remains in equilibrium with the thermal bath. When their
annihilation rate into ordinary matter becomes smaller than the
expansion of the Universe, they decouple from the thermal bath, or
``freeze-out''~\cite{kt}.  The ``freeze-out'' temperature is
typically small compared to the neutralino's mass, $T_f\sim
\mchi/20$~\cite{kt}.  Relic neutralinos are therefore always
non-relativistic, or cold, DM
candidate~\cite{ehnos,jkg}.

Many bounds on the neutralino mass and
the parameter space of SUSY models have been derived by requiring that the
neutralino abundance does not ``overclose'' the Universe. This
requires satisfying the condition $\abundchi\lsim1$~\cite{kt}, where 
$\Omega_\chi=\rhochi/\rhocrit$ and $h$ is related to the Hubble
parameter $H_0=100\,h\, \mbox{km/sec/Mpc}$. This condition comes 
from considering the evolution of a thermal population of
LSPs in the expanding Universe and in particular their annihilation
cross section at decoupling. The annihilation has to be efficient
enough to deplete the LSP number density to acceptable values.
Cosmological properties of the neutralino as the LSP and DM are often
taken into account in many studies of SUSY, including present and
future collider and DM searches.

In this Letter we will show that this standard paradigm changes
dramatically if one assumes that it is the axino, rather than the
lightest neutralino, which is the LSP. This assumption is well
justified. Experimental searches at LEP have now pushed the neutralino
mass limit considerably, above about 28\gev\ in the
MSSM\cite{lep-bound}. In more restrictive, and perhaps more motivated,
models the bound can be much higher. For example, in the Constrained
MSSM (CMSSM)~\cite{kkrw}, also known as the minimal supergravity
model, it is already around 42\gev~\cite{ellis-cosmo98}.  On the other
hand, it is worth remarking that these bounds strongly depend on the
(well-motivated) assumption that the masses of the gauginos (the fermionic
partners of the gauge bosons) are equal at a grand-unified scale. In
the absence of this condition one recovers a
robust model independent bound $\mchi\gsim3\gev$~\cite{griest:92} from
requiring $\abundchi\lsim1$ - a neutralino version of the so-called
Lee-Weinberg bound~\cite{kt}. 

In contrast to the neutralino, the mass of the axino, $\maxino$,
remains not only virtually unconstrained experimentally but also
theoretically one can easily imagine it in the few to tens of \gev\
range which we are interested in~\cite{ckn}.  This is illustrated by
the following examples. In the supersymmetric version of the heavy
quark axion model (the KSVZ model~\cite{kim}), the axino mass can
arise at one-loop level with a SUSY breaking $A$-term insertion at the
intermediate heavy squark line. Then, we expect $\maxino \sim
(f_Q^2/8\pi^2)A$ where $f_Q$ is the Yukawa coupling of the heavy quark
to a singlet field containing the axion.  In a straightforward SUSY
version of the DFSZ~\cite{dfsz} model axino mass is typically rather
small, $\maxino\sim{\cal O}(\kev)$ as has been pointed out in
Ref.~\cite{rtw}.  However, in addition to the above inevitable
contributions to the KSVZ and DFSZ axino masses, there can be other
contributions from superpotentials involving singlet fields. For
example, if the PQ symmetry is assumed to be broken by the
renormalizable superpotential term (KSVZ or DFSZ models)
$W=fZ(S_1S_2-\fa^2)$ where $f$ is a coupling, and $Z, S_1$ and $S_2$
are chiral fields with PQ charges of $0$, $+1$ and $-1$, respectively,
then the axino mass can be at the soft SUSY breaking mass scale. The
axino mass arises from the mass matrix of $\tilde S_1,\tilde S_2$, and
$\tilde Z$,

\beq \left(\matrix{0,\ m_{\axino},\ f\fa\cr m_{\axino},\ 0,\
f\fa\cr f\fa,\ f\fa,\ 0}\right).  
\eeq 

Certainly, the tree level axino mass is zero if $<Z>=0$. However, with
soft terms included, there appears a linear term in $Z$,
$V=|f|^2(|S_1|^2+|S_2|^2)|Z|^2+(AfS_1S_2Z+{\rm h.c.})$; thus $<Z>$ is
of order $A/f$, and the axino mass can arise at the soft mass scale.
A complete knowledge of the superpotential is necessary to pin down
the axino mass~\cite{ckn,cl}.  Therefore, generically it is not
unreasonable to consider the axino mass scale of order tens of GeV.

One severe bound of $\maxino\lsim2\kev$ has been derived by Rajagopal,
Turner and Wilczek\cite{rtw}. This bound arises from requiring that
the ``primordial'' axinos, produced along with the axions in the very
early Universe when the PQ symmetry becomes broken around the scale
$\fa\sim10^{11}\gev$, do not contribute too much to the total relic
abundance of the Universe, $\abunda<1$. As was noted in
Ref.~\cite{rtw}, the bound $\maxino\lsim2\kev$ (which would make the
axino a {\em warm} dark matter candidate) can be evaded by assuming
that, at temperatures below $\fa$, the Universe underwent a period of
inflation and that the temperature of subsequent 
reheating was sufficiently below
$\fa$. These assumptions are not too radical and have now become part
of the standard cosmological lore~\cite{kt}. The same way out
is also usually used to solve the analogous problem with
primordial gravitinos~\cite{ekn}.

Once the number density of the primordial axinos has been diluted by
inflation, they can be again produced in the decays of heavier
particles~\cite{kmn}. Since axino's couplings to matter are strongly
suppressed by $1/\fa$, all
heavier SUSY partners first cascade-decay to the next-to-lightest SUSY
partner (NLSP). A natural candidate for the NLSP is the lightest
neutralino. 
As stated above, the neutralino ``freezes-out'' at $T_f\sim \mchi/20$. If
it were the LSP, its co-moving number density after
freeze-out would remain nearly constant. 
In the scenario considered
here, the neutralino, after decoupling from the thermal equilibrium,
will subsequently decay into the axino via,
\eg, the process
\beq
\chi\ra\axino\gamma.
\label{chitoagamma:eq}
\eeq

This process was already
considered early on in Ref.~\cite{kmn} (see also~\cite{rtw}) in the
limit of a photino NLSP and only for both the photino and axino masses
assumed to be very low, $\mphotino \leq 1 \gev$ and $\maxino\leq 300
\ev$, the former bound now excluded by LEP searches. In that case, the
photino lifetime was typically much larger than 1 second thus normally
causing destruction of primordial deuterium produced during
nucleosynthesis by the energetic photon. Avoiding this led to a lower
$\fa$-dependent
bound on the mass of the photino in the $\mev$ range~\cite{kmn}.

In this Letter, we show that the conclusions and bounds of
Refs.~\cite{kmn,rtw} can be evaded if one considers both the axino and
the neutralino in the \gev\ mass range. In this regime the
neutralino  decays into the axino typically well before
nucleosynthesis thus avoiding the problems considered in
Refs.~\cite{kmn,rtw}. The resulting non-thermally produced axino will
be a {\em cold} DM candidate.

{\it 2. NLSP Freeze-Out}.\
The effective coupling of the neutralino with the axino
is very much weaker than that of its
interactions with other matter. Therefore the
neutralino's 
decoupling is not different from the case when it is
the LSP. The freeze--out temperature $T_f$ is determined by
the annihilation cross section $\sigma (\chi\chi \rightarrow {\rm
ordinary\ matter})$ and is normally
well-approximated by iteratively solving the equation for $\xf=T_f/\mchi$
\beq
{\frac{1}{\xf}}=\ln\left[{\frac{\mchi}{4\pi^3}}\mplanck
\sqrt{\frac{45\xf}{N_F}}
\VEV{\sigma v_{rel}}(\xf)\right],
\label{inversexf:eq}
\eeq
where $\mplanck= 1.22\times10^{19}\gev$, $N_F$ is the
effective number of relativistic degrees of freedom and $\langle
\sigma v_{rel} \rangle$ is the averaged product of
annihilation cross section and the annihilating
neutralinos' relative velocity. As already mentioned, 
the scaled freeze-out temperature $\xf$ 
is typically very small ($\xf={\cal O}(1/20)$), justifying
the iterative procedure~\cite{kt}. 
Without further decay, the neutralino co-moving number
density $\nchi$ would have remained basically constant.

\begin{figure}[htbp]
\centerline{\psfig{file=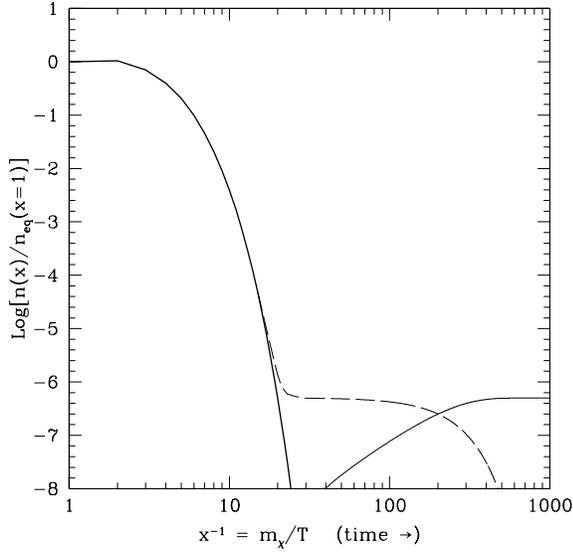,width=3in}}
\bigskip
\caption{
A schematic behavior of the co-moving number density: the thermal
equilibrium (thick solid), NLSP neutralino (dash) and LSP axino (thin
solid). 
}
\label{freezeout:fig}
\end{figure}

\noindent
{\it 3. Neutralino decay into axino}.
In the scenario considered here, the NLSP neutralino 
co-moving number density after ``freeze-out'' will continue
to decrease because of its 
decay~(\ref{chitoagamma:eq}) into the axino
LSP. This is presented in Fig.~\ref{freezeout:fig}.
At  $T/\mchi\equiv x<\xf$ and 
for $\Gamma_\chi \ll H$,  $\nchi$ is roughly given by
\beq
\nchi (x) \simeq {\nchi}^{eq} (\xf) C(x)  
\exp\[-\int_{x}^{x_f} \frac{dx'}{x'^3}{\langle \Gamma_\chi\rangle_{x'} \over
H(m_\chi) }\]
\label{nchidecay:eq}
\eeq
where $C(x)$ takes into account the temperature difference between
the photons and the decoupled neutralinos and 
$\langle \Gamma_\chi \rangle_x $ is the 
thermally averaged decay rate for the neutralino at $x$, while
$ H(m_\chi) = 2\pi \sqrt{2\pi N_F/45}\  m^2_\chi/M_P$.

The interaction of the axino and the gaugino component 
of the neutralino is given by the term 
$\alpha_Y\cayy/(4\sqrt{2}\pi\fa) [(\Phi B_\alpha
B^\alpha)_{\theta\theta} + (\Phi^\ast {\bar B}_{\dot\alpha} {\bar
B}^{\dot\alpha})_{{\bar\theta}{\bar\theta}} ] +$
$\alpha_2\caww/(4\sqrt{2}\pi\fa)
[B \ra W_3 ]$, where $\Phi$ is the chiral supermultiplet containing the
axion and the axino, while the vector multiplet $B$ ($W_3$)
corresponds to the $U(1)_Y$ ($SU(2)_L$) gauge group with a coupling
strength $\alpha_Y$ ($\alpha_2$). The coefficients
$\cayy$ and $\caww$ are model dependent.  Usually, one performs chiral
transformations so that there is no axion--$W_{\mu\nu}\tilde
W^{\mu\nu}$ coupling. This is equivalent to giving vanishing
Peccei-Quinn charges to left-handed doublets.  In this case,
$C_{aWW}=0$ and $C_{aYY}= C_{a\gamma\gamma}$.  In the DFSZ model with
$(d^c,e)$--unification $C_{aYY}=8/3$, and in the KSVZ model for
$e_Q=0,-1/3,$ and 2/3 $C_{aYY}=0,2/3$ and 8/3,
respectively~\cite{coupling}. Below the QCD chiral symmetry breaking
scale, $C_{a\gamma\gamma}$ and $C_{aYY}$ are reduced by $1.92$.

We first concentrate on the dominant decay channel~(\ref{chitoagamma:eq})
which is always allowed as long as $\maxino<\mchi$.
We will comment on other channels below.
The decay rate for the process~(\ref{chitoagamma:eq}) is given by
\beq
\Gamma = {\alpha^2_{em} N^2 \over 16 \pi^3} \cachigamma^2
{{\mchi}^3\over \fa^2} 
\(1-{m^2_{\tilde a}\over m^2_\chi}\)^3,
\label{gammachi:eq}
\eeq
where $\alpha_{em}$ is the electromagnetic coupling strength,
$\cachigamma=(\cayy/\cos\theta_W) Z_{11}$, and 
$N$ is a model dependent factor ($N=1(6)$ for the KSVZ (DFSZ) model).

In the theoretically most favored case of a
nearly pure B-ino~\cite{chiasdm,kkrw}, 
the neutralino lifetime can be written as
\beq
\tau\simeq 3.3 \times 10^{-2} {\rm sec}\, {\frac{1}{\cayy^2}}
\left(\frac{f_a/N}{10^{11}\gev}\right)^2
\left(\frac{50\gev}{\mchi}\right)^3
\label{binolife:eq}
\eeq
where the phase 
space factors from Eq.~(\ref{gammachi:eq}) have been neglected.

A comment is in order regarding a plausible range of
$\fa$~\cite{raffelt-rev}. A somewhat model dependent lower bound
$\fa\gsim10^9\gev$ comes from astrophysical considerations, most
notably from requiring that axions do not overly affect processes in
globular clusters, red giants and in supernova 1987A. An upper limit
of $\sim10^{12}\gev$ is quoted in the context of cold axion energy
density.  A range $10^{9-10}\gev\lsim\fa\lsim10^{12}\gev$ gives a
cosmologically interesting values for the relic density of axions.

Hard photons produced in the $\chi$ decay thermalize via multiple
scattering from background electrons and positrons
($\gamma+e\ra\gamma+\gamma+e$)~\cite{dkt,kmn}. The process proceeds
rapidly for electromagnetic background temperatures above $1\mev$
at which point background $e-\bar e$ pairs annihilate. To ensure efficient
thermalization and in order to avoid problems with photodestruction of
light elements produced during nucleosynthesis, we require that the
neutralino lifetime~(\ref{binolife:eq}) is sufficiently less than about
1 second (which also coincides with temperatures of about $1\mev$). A
modest requirement $\tau\lsim10^{-1}{\rm sec}$ leads, in the case of
the neutralino with a large B-ino component (a neutralino is never a
{\em pure} B-ino state), to an upper bound on $\fa$ which depends on
$\mchi$. Also, at larger
masses additional decay channels open up, most notably $\chi\ra
Z\axino$.  We can see that, for large enough 
values of the B-ino mass, the decay~(\ref{chitoagamma:eq}) will almost
entirely take place before nucleosynthesis.

The case of higgsino-dominated neutralino as the NLSP is probably less
attractive and also more model dependent. First, the
lifetime~(\ref{binolife:eq}) is now be typically significantly larger,
easily extending into the period of nucleosynthesis and beyond. This
is caused by the suppression of the B-ino component through which the
decay proceeds. (See the form of $\cachigamma$ below
Eq.~(\ref{gammachi:eq})).  Much lower values of $\fa$ could be
considered as a remedy or much larger higgsino masses, and/or
additional (model dependent) decay channels involving Higgs in the
final state.

In the MSSM, the higgsino relic abundance in the mass range allowed by
LEP is typically very small, thus leading to even smaller
$\abunda$. One possibility would be to consider rather obese higgsino
masses, above roughly 500\gev, where $\abundchi\gsim1$ again. A
resulting value of $\abunda$ would then depend on the actual size of
the higgsino component in the decaying neutralino, as well as on the
axion/axino model which would determine the couplings of the decay
channels to the Higgs final state. One could also allow for a Higgs
singlet and assume that its fermionic partner is mostly the NLSP. 

The resulting axino relic abundance today is simply given by 
\beq
\abunda= {\maxino\over\mchi}\abundchi
\label{abunda:eq}
\eeq
since all the neutralinos have  decayed into axinos. The axino will
normally be produced relativistic, except when the 
ratio of the neutralino-axino mass difference to the axino mass is small, 
but will later 
redshift due to the expansion of the Universe 
and become cold by the time of matter dominance.
It is worth noting that the neutralinos will not dominate the energy 
density of the Universe before decaying; to see this we have to compare
its lifetime, given by Eq.~(\ref{binolife:eq}), with the time when
the equality $\rho_\chi = \rho_{rel}$ takes place. This
time is easily computed neglecting the decay and amounts
to $10^6-10^7 {\rm sec}$; we see therefore that the neutralinos
never dominate the energy density and matter domination
starts only after the produced axinos become non-relativistic.

With the lifetime~(\ref{binolife:eq}) significantly larger than
$10^{-6} {\rm sec}$ the neutralinos escape the high energy
detectors. Thus phenomenology remains basically unchanged from the
usual case where the neutralino is the LSP. In particular, accelerator
mass bounds on supersymmetric particles apply. But cases previously
excluded by the constraint $\abundchi<1$ can now be allowed via
Eq.~(\ref{abunda:eq}). This leads to a possibly dramatic relaxation of
the parameter space of SUSY models. For example, in the MSSM the
region of large higgsino masses mentioned above has been considered
cosmologically excluded but now can be again allowed if one takes a
sufficiently small ratio $\maxino/\mchi$. In the gaugino region it is
normally reasonable to expect that, in order to reduce the LSP
relic abundance below one~\cite{chiasdm}, at least one sfermion with
mass roughly below  500\gev\ should exist.  In the CMSSM, the same
requirement often provides an upper bound on the common scalar and
gaugino masses of order one \tev~\cite{kkrw} over a large range of
parameters. Both bounds which hold for a gaugino-like LSP away from
annihilation resonances can now be readily relaxed.

So far we have considered the neutralino as the NLSP. This assumption
can easily be relaxed to accommodate any other superpartner, either
neutral or carrying an electric or color charge, provided that its
effective coupling with the axino is of order $\sim1/\fa$. 
All one needs to require is that the NLSP decay into the axino and the
accompanying ordinary particle thermalization takes place before
nucleosynthesis. If this can be achieved then cases previously
considered excluded as corresponding to non-neutral LSPs can now again
be allowed.

In conclusion, we have shown that the axino can easily be the LSP with
a mass in the $\gev$\ range. Such an axino would be a {\em cold} dark
matter candidate for a natural range of the Peccei-Quinn scale
$\fa$. It is not impossible that, with or without its
non-supersymmetric partner, the axino could dominate the
matter in the Universe.

\acknowledgments
One of us (JEK) is supported in part by Korea Science and Engineering
Foundation, Ministry of Education through
BSRI 98-2468, and Korea Research Foundation. LR would like to
acknowledge kind hospitality of KIAS (Korea Institute for Advanced
Study) where part of the project was done, and to thank
A.~Masiero for helpful conversations.


\begin{thebibliography}{99}

\bibitem{pq} R. D.~Peccei and H. R.~Quinn, Phys. Rev. Lett. {\bf 38},
1440 (1977); Phys. Rev. {\bf D16}, 1791 (1977).

\bibitem{axion} S.~Weinberg, Phys. Rev. Lett. {\bf 40}, 223 (1978);
F.~Wilczek, Phys. Rev. Lett. {\bf 40}, 279 (1978).

\bibitem{kim} J. E.~Kim, Phys. Rev. Lett. {\bf 43}, 103 (1979); M. A.~Shifman,
V. I.~Vainstein, and V.I.~Zakharov, Nucl. Phys. {\bf B166}, 4933 (1980).

\bibitem{dfsz} 
M. Dine, W. Fischler, and M. Srednicki, Phys. Lett. {\bf 104B},
99 (1981); A.~P.~Zhitnitskii, Sov. J. Nucl. Phys. {\bf 31}, 260 (1980).

\bibitem{kt} E.~W.~Kolb and M.~S.~Turner, {\it The Early Universe}
(Addison-Wesley, Redwood City, 1990).

\bibitem{ehnos} J.~Ellis, J.~S.~Hagelin, D.~V.~Nanopoulos, K.~Olive, and
M.~Srednicki, Nucl. Phys. {\bf B238}, 453 (1984).

\bibitem{jkg} For a review, see G.~Jungman,
M.~Kamionkowski, and K.~Griest, Phys. Rep. {\bf 267}, 195 (1996).

\bibitem{lep-bound} For a recent review, see, S.~Katsanevas, talk at
the 2nd International Workshop on the Identification of Dark Matter
(IDM-98), Sheffield, England, September, 7 - 11, 1998.

\bibitem{kkrw} G.~L.~Kane, C.~Kolda, L.~Roszkowski, and J.~Wells,
Phys. Rev. {\bf D49}, 6173 (1994); R.~G.~Roberts and L.~Roszkowski,
Phys. Lett. {\bf B309}, 329 (1993).

\bibitem{griest:92} K.~Griest and L.~Roszkowski,
Phys. Rev. {\bf D46}, 3309 (1992).

\bibitem{ellis-cosmo98} J.~Ellis, {\it et al.}, Phys. Rev. {\bf D58},
095002 (1998).

\bibitem{ckn} E.~J. Chun, J.~E. Kim, and H.~P. Nilles, Phys. Lett {\bf B287},
123 (1992).

\bibitem{rtw} K.~Rajagopal, M.~S.~Turner, and F.~Wilczek, Nucl. Phys. {\bf
B358}, 447 (1991).


\bibitem{cl} E.~J. Chun and A. Lukas, Phys. Lett. {\bf B357}, 43 (1995).

\bibitem{ekn} J. Ellis, J.~E. Kim, and D.~V. Nanopoulos,
Phys. Lett. {\bf B145}, 181 (1984).

\bibitem{kmn} J.~E. Kim, A. Masiero, and D.~V. Nanopoulos, Phys. Lett.
{\bf B139}, 346 (1984).

\bibitem{coupling} J.~E.~Kim, Phys. Rev. {\bf D58}, 055006 (1998).
See, also, D.~B.~Kaplan, Nucl. Phys. {\bf B260}, 215 (1985);
M.~Srednicki, {\it ibid.} {\bf B260}, 689 (1985).

\bibitem{chiasdm} L.~Roszkowski, Phys. Lett. {\bf B262}, 59 (1991).

\bibitem{raffelt-rev} 
See, 
\eg, G.G.~Raffelt,
hep-ph/9806506.


\bibitem{dkt} D.~A.~Dicus, E.~W.~Kolb, and V.~L.~Teplitz,
Astrophys. J. {\bf 221}, 327 (1979).

\end{thebibliography}
\end{document}